\documentclass[sn-mathphys]{sn-jnl}
\jyear{2021}%
\usepackage{algorithm}
\usepackage{algpseudocode}
\theoremstyle{thmstyleone}%
\theoremstyle{thmstyletwo}%

\theoremstyle{thmstylethree}%

\begin{document}

\title[Compressed Out-of-Core Stencil]{Compression-Based Optimizations for Out-of-Core GPU Stencil Computation}

\author*[1]{\sur{Jingcheng Shen}}\email{shenjc@cqupt.edu.cn}

\author[1]{\sur{Xin Deng}}\email{dengx@cqupt.edu.cn}

\author[2]{\sur{Yifan Wu}}\email{yf-wu@ist.osaka-u.ac.jp}

\author[2]{\sur{Masao Okita}}\email{okita@ist.osaka-u.ac.jp}

\author[2]{\sur{Fumihiko Ino}}\email{ino@ist.osaka-u.ac.jp}

\affil[1]{\orgdiv{School of Computer Science and Technology}, \orgname{Chongqing University of Posts and Communications}, \orgaddress{\street{2 Chongwen Rd.}, \city{Nan'an Dist.}, \postcode{400065}, \state{Chongqing}, \country{China}}}

\affil[2]{\orgdiv{Graduate School of Information Science and Technology}, \orgname{Osaka University}, \orgaddress{\street{1-5 Yamadaoka}, \city{Suita}, \postcode{565-0871}, \state{Osaka}, \country{Japan}}}

\abstract{An out-of-core stencil computation code handles large data whose size is beyond the capacity of GPU memory. Whereas, such an code requires streaming data to and from the GPU frequently. As a result, data movement between the CPU and GPU usually limits the performance. In this work, compression-based optimizations are proposed. First, an on-the-fly compression technique is applied to an out-of-core stencil code, reducing the CPU-GPU memory copy. Secondly, a single working buffer technique is used to reduce GPU memory consumption. Experimental results show that the stencil code using the proposed techniques achieved 1.1$\times$ speed and reduced GPU memory consumption by 33.0\% on an NVIDIA Tesla V100 GPU.}




\maketitle

\section{Introduction}\label{sec1}
Stencil computation is an important class of scientific application which pertains in a wide range of research and industry fields, such as geophysics simulations~\cite{serpa17padw,farres19eage,shen20ieice}, computational electromagnetics~\cite{adams07fdtd}, and image processing~\cite{tabik18supe}.
Stencil computation updates every element in given arrays (i.e., datasets) according to one or more fixed calculating patterns (i.e., \texttt{stencil}s), which is an embarrassingly parallelizable task for graphics processing units (GPUs).
A GPU has thousands of cores and its memory bandwidth is 5--10 times higher than that of a CPU, thus excelling at accelerating both compute- and memory-intensive scientific applications~\cite{shen19ccpe,nogueira2019accelerating,garcia2022parallel}. 
However, as a GPU has a limited capacity of device memory (tens of GBs), it fails to directly run a large stencil code whose data size exceeds the memory capacity.

A large entity of research on GPU-based out-of-core stencil computation has been performed to address this issue~\cite{jin14ichpca,sourouri16ijpp,shimokawabe17iccc,miki19hpcn,shen20ieice}. For a large dataset whose data size exceeds the capacity of the device memory, out-of-core computation first decomposes the dataset into smaller blocks and then streams the blocks to and from the GPU to process. Nevertheless, the performance of this approach is often limited by data movement between the CPU and GPU because the interconnects fail to catch up with the development of the computation capability of GPUs as described in \cite{shen19ccpe}. 

Existing techniques such as temporal blocking and region sharing to reuse the on-GPU data and to avoid redundant data transfer \cite{jin14ichpca,miki19hpcn,shen20ieice} were studied, yet new optimizations are in demand to further reduce data movement overhead.

On-the-fly compression based techniques are promising. Such techniques can be used to compress the data on the GPU before moving it to the CPU and decompress the data on the GPU before processing. Studies on the acceleration of GPU-based out-of-core stencil computation with on-the-fly compression are rare. According to a comprehensive review \cite{cappello20review}, studies on leveraging compression techniques in scientific applications mainly focused on scenarios such as post-analysis and failure recovery. 

To fill the research gap, this article proposes compression-based optimization techniques for out-of-core stencil computation. The major contribution of this article is therefore summarized as follows:
\begin{itemize}
    \item An on-the-fly compression technique is integrated into an out-of-core stencil code to reduce the amount of data movement between the CPU and GPU, shifting the performance bottleneck from data movement to GPU computation.
    \item A single working buffer technique is proposed, which significantly reduces GPU memory consumption.
    \item A detailed analysis is given to demonstrate the usefulness of our approach to optimize out-of-core GPU stencil computation.
\end{itemize}

The remainder of this article is organized as follows: Related work on accelerating stencil and similar scientific applications with compression techniques are introduced in Section~\ref{sec2}. Background of out-of-core stencil computation are briefly described in Section~\ref{sec3}. Section~\ref{sec4} gives the basics of integrating an on-the-fly compression technique into out-of-core computation. The proposed memory-saving technique is described in Section~\ref{sec5}. In Section~\ref{sec6}, experimental results are presented and analyzed. Finally, Section~\ref{sec7} concludes the present work and suggests future research directions.

\section{Related work}\label{sec2}

Tao \textit{et al.}~\cite{tao18hpdc} and Calhoun \textit{et al.}~\cite{calhoun19saga} used lossy compression techniques to compress checkpointing data to improve restart performance. 
Jin \textit{et al.}~\cite{jin20ipdps} used GPU-based lossy compression for post-analysis for cosmological simulations.
These studies are more relevant to a stand-alone procedure rather than an on-the-fly one.

Wu \textit{et al.}~\cite{wu19sc} simulated large quantum circuits using lossy or/and lossless compression techniques adaptively, increasing the simulation size by 2--16 qubits. Their research is based on a CPU-based supercomputer.

Zhou \textit{et al.}~\cite{zhou21ipdps} extended MPI libraries to reduce communication time in a GPU cluster by compressing the messages transferred between nodes. The size of messages was up to 32 MB. On the other hand, our method compressed large datasets for stencil computation that were more than 10 GB to reduce the data transfer time between the CPU and GPU.

Sun \textit{et al.}~\cite{sun22trets} proposed an accelerator platform that eliminates the data movement bottleneck between PCIe-attached FPGAs and their host servers via compression.
Their approach mainly focuses on optimizing the ZFP compression algorithm \cite{cuZfp} on a hardware (i.e. FPGA) level. On the other hand, our work does not change the compression algorithm itself, but focuses on how to efficiently use compression in the scenario of GPU acceleration.

\section{Background}\label{sec3}
\begin{figure}[tb!]
    \centering
    \includegraphics[width=.85\hsize]{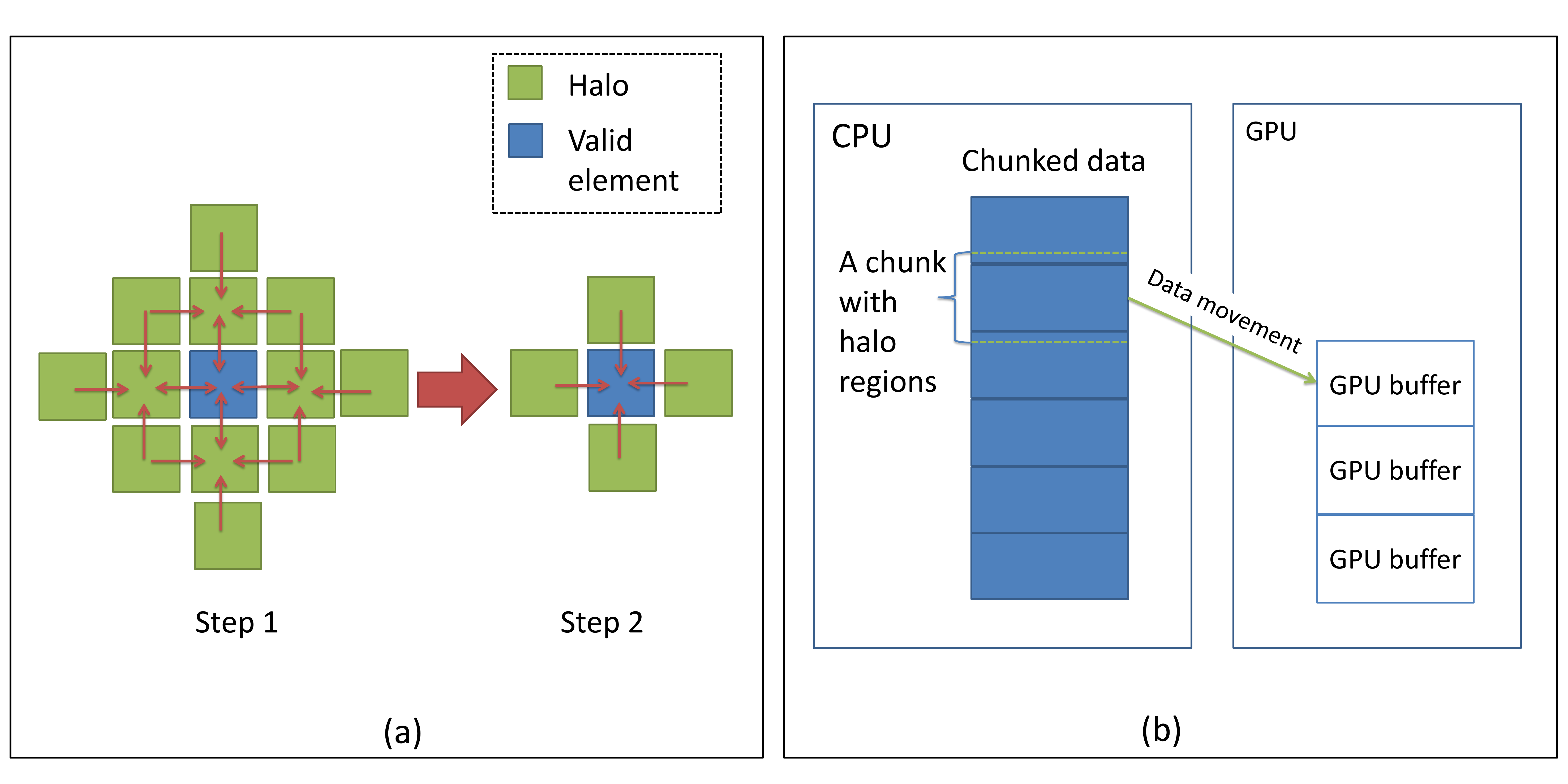}
    \caption{Example of out-of-core stencil computation. (a): Updating an element in a five-point stencil computation code for two time steps. (b) Transferring data chunks with halo regions.}
    \label{fig:stencil}
\end{figure}

\begin{figure}[tb!]
    \centering
    \includegraphics[width=.9\hsize]{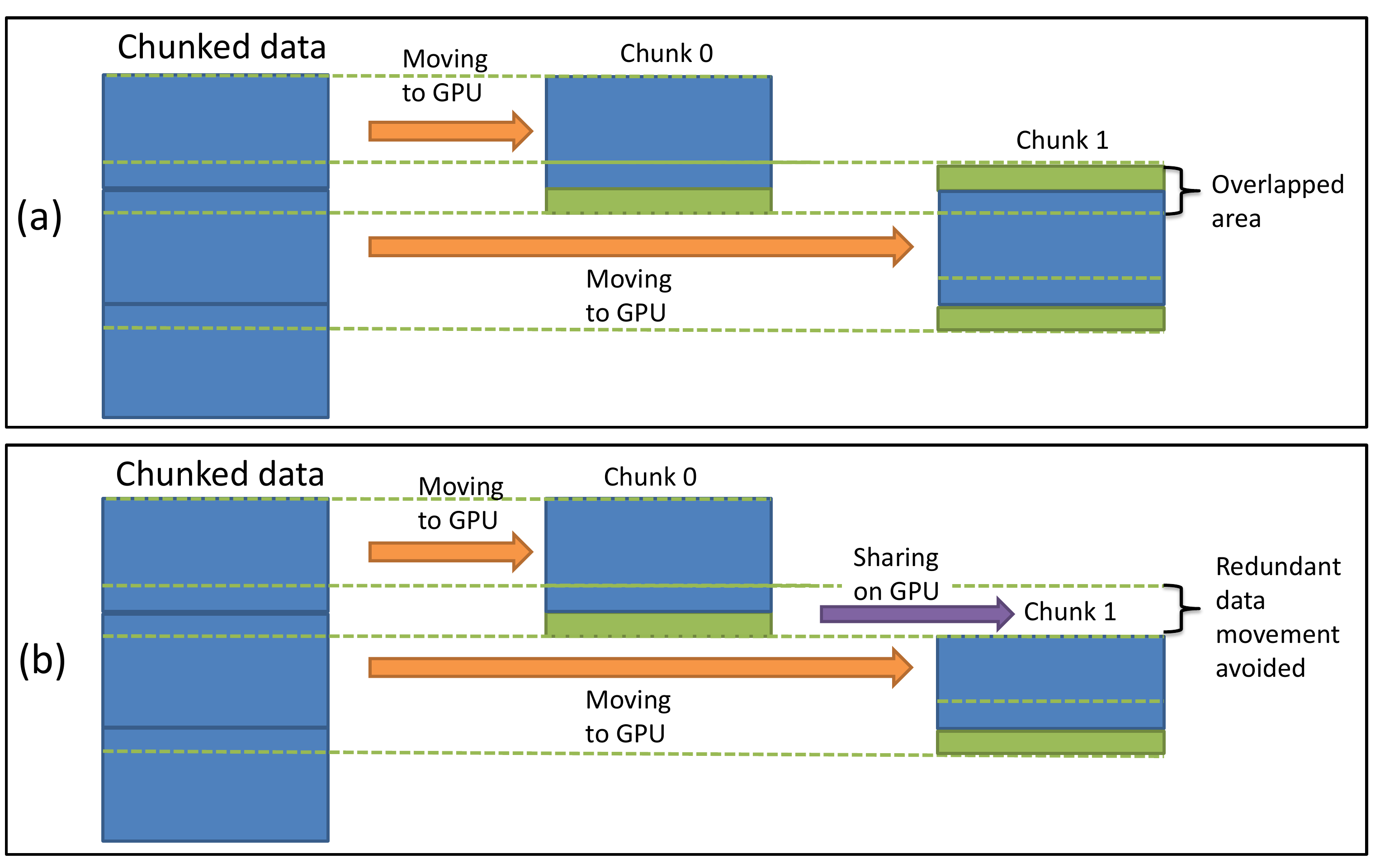}
    \caption{Region sharing. (a): Two contiguous chunks have an overlapped area. (b): Avoiding redundant data movement by allowing two contiguous chunks to share the overlapped area on the GPU.}
    \label{fig:rs}
\end{figure}

The background of out-of-core stencil computation is provided in this section. Above all, we explain the reason why out-of-core stencil computation can benefit from compression.

\subsection{Out-of-core stencil computation}\label{subsec31}
Stencil computation is an iterative data processing solution that updates array elements according to one or more fixed patterns called \texttt{stencils}. In a stencil, the value of the element to be updated is computed with the values of surrounding elements called \texttt{halo} (Fig.~\ref{fig:stencil}(a)).

Out-of-core approaches tackle with large data of stencil computation that is beyond the capacity of GPU memory. Out-of-core stencil computation decomposes the original data into smaller data chunks, each of which can fit in the GPU memory. Chunks are streamed to and from the GPU for processing. However, such an approach incurs frequent data movement between the CPU and GPU, which is prone to performance degradation. 

\texttt{Temporal blocking} \cite{miki19hpcn}, is a widely used optimization to reduce data movement in hierarchical memory systems. In the scenario of out-of-core stencil computation, temporal blocking piggybacks halo regions with chunks according to the number of time steps we want to compute with chunks on the GPU (Fig.~\ref{fig:stencil}(b)). In doing so, we can reduce the time of CPU-GPU data movement.

A \texttt{region sharing} technique \cite{jin14ichpca,reguly17mchpc,shen20ieice} can be leveraged to further reduce CPU-GPU data movement. That is, two contiguous chunks share a common (i.e., overlapped) area (Fig.~\ref{fig:rs}(a)), and a chunk moved to the GPU can therefore share such common data on the GPU to the next chunk. As a result, we can move the next chunk to the GPU without the redundant data (Fig.~\ref{fig:rs}(b)).

\subsection{Data movement bottleneck}\label{subsec32}
\begin{figure}[tb!]
    \centering
    \includegraphics[width=.65\hsize]{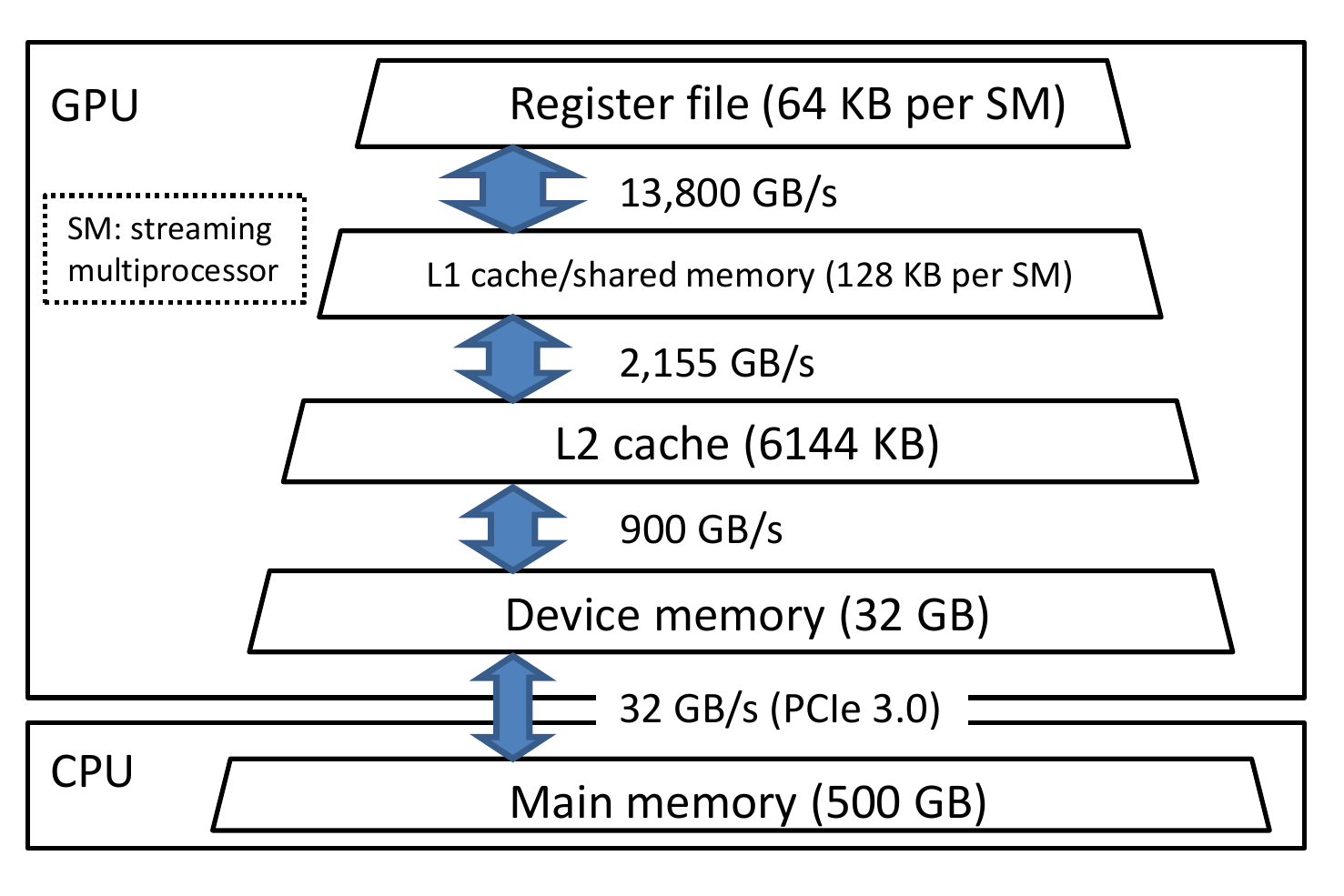}
    \caption{Hierarchical memory system of the testbed in the work that comprises an NVIDIA Tesla V100 GPU \cite{wwwv100,jia18} and an Intel Xeon Silver 4114 CPU. The PCIe 3.0 interconnect between the CPU main memory and GPU device memory limits the performance of the system in terms of data movement.}
    \label{fig:bandw}
\end{figure}

\begin{figure}[tb!]
    \centering
    \includegraphics[width=\hsize]{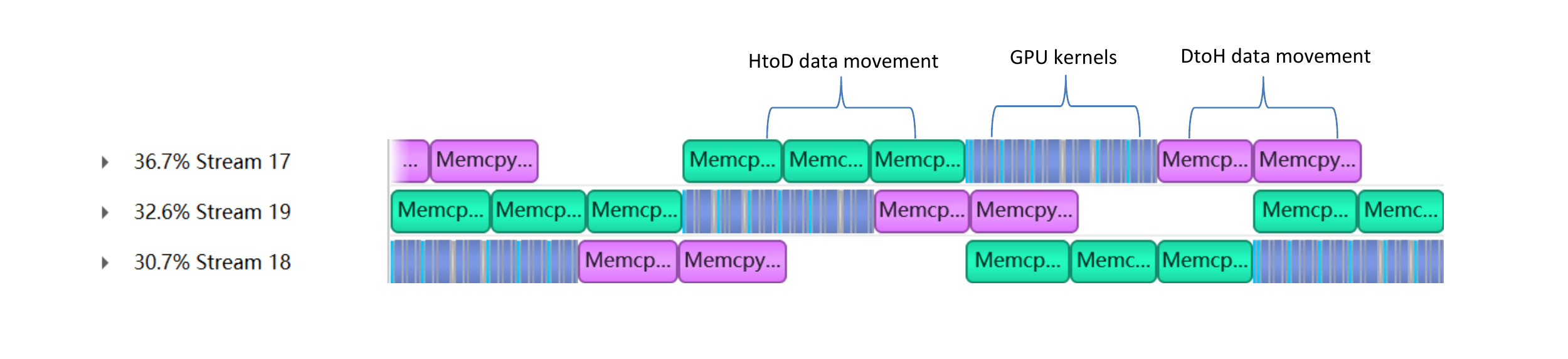}
    \caption{A part of output of profiling an out-of-core GPU stencil code with NVIDIA Nsight Systems \cite{nsys22}. CPU-to-GPU (HtoD) data movement, GPU kernels, and GPU-to-CPU (DtoH) movement are overlapped with each other using three CUDA streams. Note that (1) the overhead of data movement is longer than GPU kernel time, becoming the performance bottleneck, and (2) GPU kernels are not mutually overlapped.}
    \label{fig:pipe1}
\end{figure}

Temporal blocking requires moving valid data with halo data to the GPU. The more time steps you want to compute using a data chunk on the GPU, the more halo data the chunk must piggyback. Throughout the hierarchical structure of CPU-GPU memory system (Fig.~\ref{fig:bandw}), the interconnect between the CPU memory and the GPU memory has a relatively small bandwidth, making the performance of an out-of-core GPU application sensitive to increased CPU-GPU data movement. 
Figure~\ref{fig:pipe1} shows that the CPU-GPU data movement limited the performance of a 25-point out-of-core GPU stencil code \cite{shen20ieice}, even though the code was implemented the aforementioned techniques, which calls for techniques to further reduce the data movement.

\section{Integrating on-the-fly compression}\label{sec4}
This section provides the basics of integrating on-the-fly compression into out-of-core stencil computation. We minimize the length of this section because you can refer to our previous work~\cite{shen22pdcat} for more details.

\subsection{Separate compression}\label{subsec41}
As mentioned in Section~\ref{subsec31}, contiguous chunks share an overlapped area on the GPU to avoid redundant data transfer. Therefore, we compress the overlapped area of a chunk separately from and the remnant of the chunk, in order to solve runtime data dependency.

\subsection{Pipeline execution}\label{subsec42}
As shown in Fig.~\ref{fig:pipe1}, CUDA streams are used to overlap data movement with computation, to some extent hiding the overhead of data movement.

Our approach utilizes a powerful GPU-based compression library cuZFP~\cite{cuZfp}, which excels at both execution time and information preservation with a customized compression rate. We modified the original project of cuZFP to conform to pipeline execution, overlapping CPU-GPU data movement with GPU kernels including decompression, computation, and compression.

\section{Single working buffer technique}
\label{sec5}

\begin{figure}[tb!]
    \centering
    \includegraphics[width=.8\hsize]{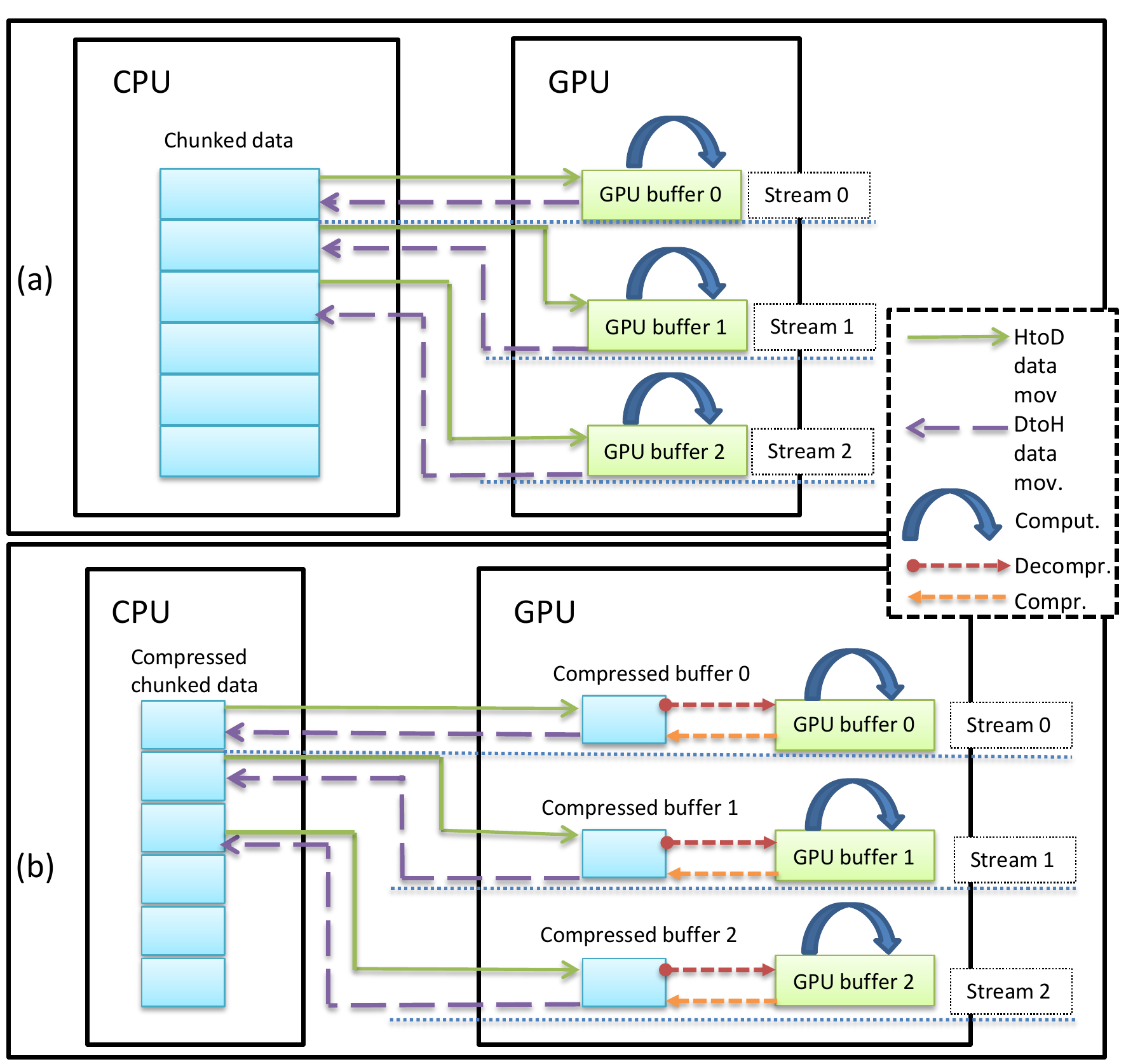}
    \caption{Architecture of an out-of-core stencil program (a) w/o compression or (b) w/ compression. Three CUDA streams are used to overlap data movement and GPU computation.}
    \label{fig:3ver}
\end{figure}

\begin{figure}[tb!]
    \centering
    \includegraphics[width=0.9\hsize]{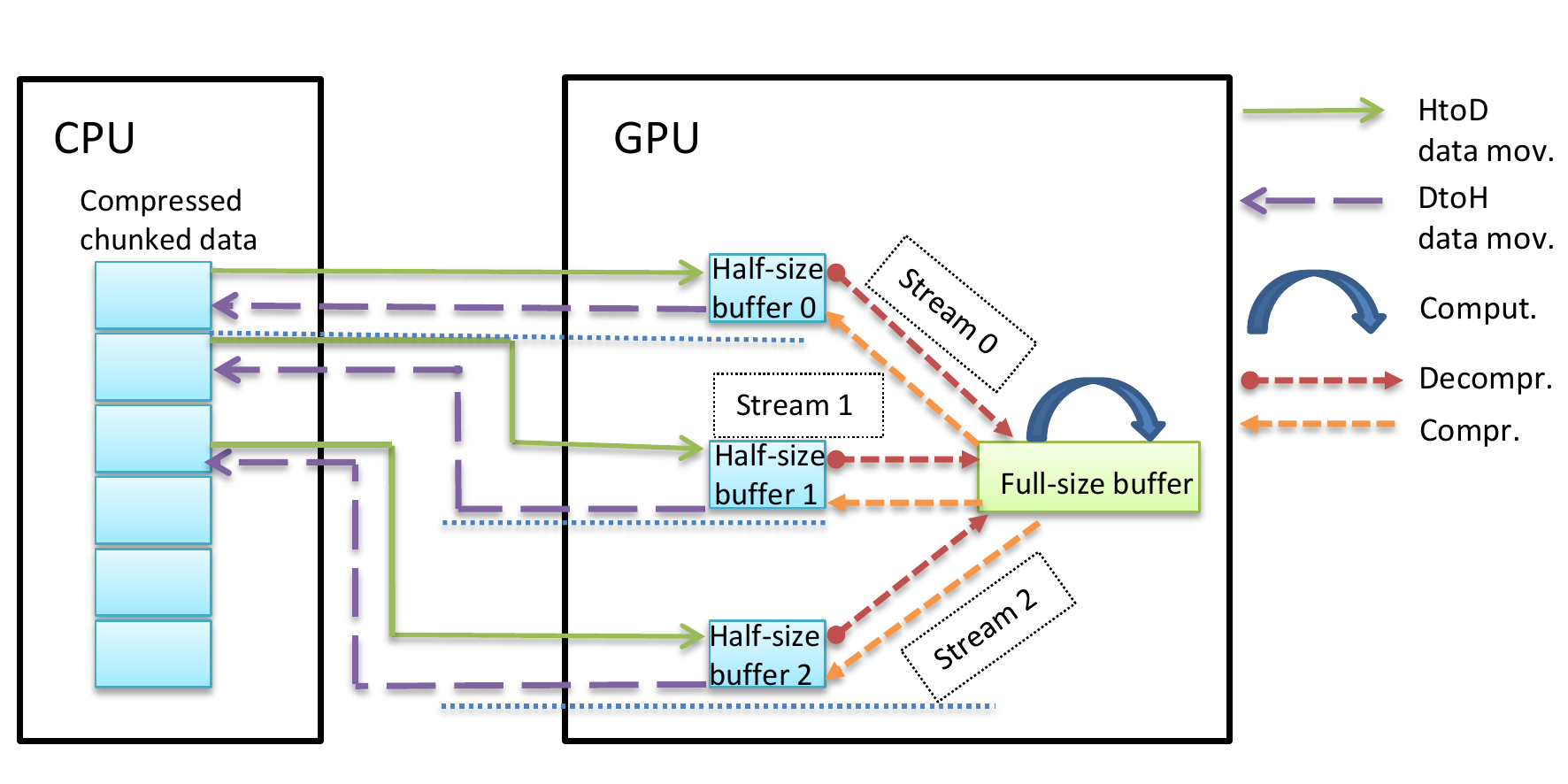}
    \caption{An out-of-core stencil program w/ compression and proposed single working buffer method. Note that we still have three compressed (i.e. half-size) buffers yet only one full-size working buffer.}
    \label{fig:swb}
\end{figure}

\begin{figure}[tb!]
    \centering
    \includegraphics[width=0.8\hsize]{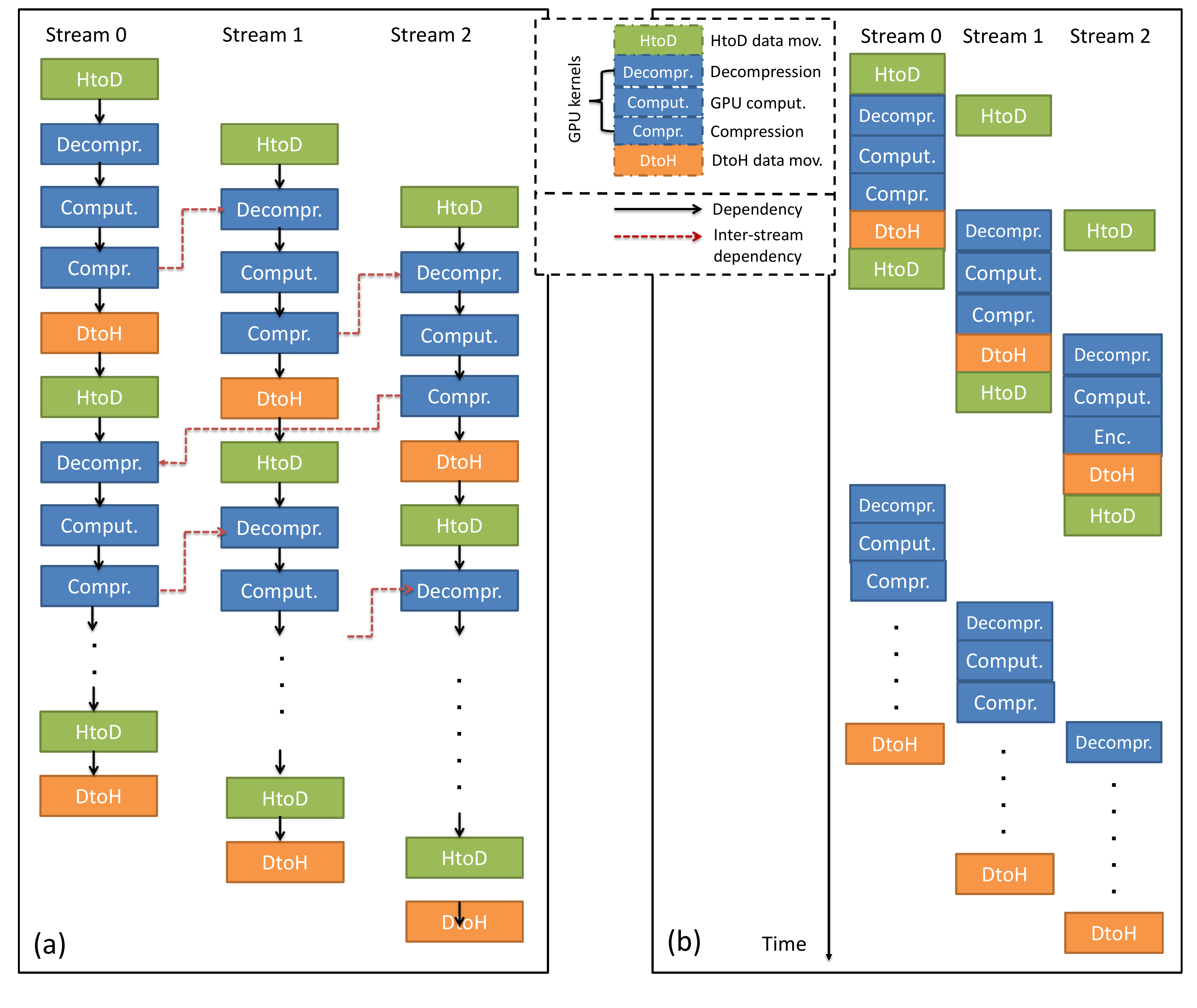}
    \caption{Securing correct execution order for proposed single working buffer method: (a) directed acyclic graph (DAG) showing dependencies among operations of the stencil program using compression and single working buffer and (b) scheduling operations to avoid resource conflicts.}
    \label{fig:dag}
\end{figure}

\begin{algorithm}[tb!]
\caption{Conflict-free scheduling.}
\label{algo:tpsort}
\begin{algorithmic}
\Require (1) $chk$[0:$n$], compressed data chunks on the CPU, (2) $strm$[0:3], CUDA streams, (3) $evt$[0:3], CUDA events, (4) $hf\_buf$[0:3], half-size buffers on the GPU to manage compressed chunk, (5) $fl\_buf$, full-size working buffer on the GPU to perform kernels
\Ensure Updated $chk$[0:$n$]
\State $si \gets 0$
\State $prev_s \gets -1$ \Comment{Index of previous stream}
\State $prev_c \gets -1$ \Comment{Index of previous chunk}
\For{$i = 0$ to $n-1$}
\If{$prev\_s \neq -1$}
    \State Compress data in $fl\_buf$ to $hf\_buf$[$prev_s$] using $strm$[$prev_s$]
    \State Record $evt$[$prev_s$] using $strm$[$prev_s$]   
    \State Transfer data in $hf\_buf$[$prev_s$] to $chks$[$prev_c$] using $strm$[$prev_s$]
    \State $prev_s \gets -1$
    \State $prev_c \gets -1$
\EndIf
\State Transfer data in $chk$[$i$] to  $hf\_buf$[$si$] using $strm$[$si$]
\State $prev \gets (si - 1) < 0$ ? $2 : (si - 1) $ 
\State Wait $evt$[$prev$] using $strm$[$si$]
\State Decompress data in $hf\_buf$[$si$] to $fl\_buf$ using $strm$[$si$]
\State Compute on data in $fl\_buf$ using $strm$[$si$]
\State $prev_s \gets si$
\State $prev_c \gets i$
\EndFor
\end{algorithmic}
\end{algorithm}

In a GPU stencil code without compression, a three-stage pipeline is most commonly used to overlap CPU-GPU data movement with GPU computation. Implementing such an approach with the CUDA language requires three CUDA streams and each stream needs a GPU buffer to store the data for GPU computation (Fig.~\ref{fig:3ver}(a)). In our previous work, we also use three CUDA streams, each of which requires a compressed buffer and a full-size working buffer. The size of a compressed buffer is determined by a user-specified compression rate. We use 32 bits to store a double floating-point value, making the compression rate 32/64$=$1/2 (i.e. half size). Compressed data is first moved from the CPU to the half-size buffers, and is then decompressed to the full-size buffer for computation (Fig.~\ref{fig:3ver}(b)). Such an approach is straightforward to implement but increases GPU memory consumption compared to a stencil code without compression. 

To improve the efficiency of memory usage, we propose a single working buffer method. Observing that GPU kernels cannot overlap with each other for stencil computation with large datasets (Fig.~\ref{fig:pipe1}), we determine that using three full-size working buffers is unnecessary. Therefore, we reduce the number of working buffers to one. 
Each of the three half-size buffers stores its own data assignment, and in turn uses the full-size working buffer to perform decompression, computation, and compression (Fig.~\ref{fig:swb}). Nevertheless, although such a method reduces the amount of GPU memory consumption, we must carefully orchestrate the three CUDA streams to avoid resource conflicts because now the streams share the same working buffer.

Task graph based methods \cite{huang21tpds,huang21tcad} can be used to schedule the GPU kernels to prevent resource conflicts, considering inter-stream dependencies. Figure~\ref{fig:dag}(a) illustrates the directed acyclic graph (DAG) of an out-of-core GPU stencil computation code using compression and the proposed single working buffer method. 
An arrowed line across streams denotes an inter-stream dependency that a stream waits for its cyclically prior stream to take control of the working buffer to execute GPU kernels (decompression, computation, and compression).

We can schedule the operations by applying topological sorting \cite{kahn62acmc} to the DAG (Fig.~\ref{fig:dag}(b)). CUDA events~\cite{eventuse} are used to realize fine-grained synchronizations between streams. At the beginning of an arrowed dotted line, we record an event with the stream the line starts from, whereas at the end of the arrowed dotted line, we wait for the event with the stream the arrow is pointed to. Algorithm~\ref{algo:tpsort} shows the proposed method of scheduling GPU kernels without resource conflicts.

\section{Experimental results}
\label{sec6}

\begin{table}[tb!]
    \centering
    \caption{Target stencil code.}
    \begin{tabular}{lllll} \\\hline
        No. of & Data type & Dim. info. & Entire   \\
        datasets &  &  & data size   \\\hline
        4 & Double & (1152+2$\times$HALO)$^3$, HALO$=$4 &  46 GB  \\\hline
    \end{tabular}
    \label{tab:dataset}
\end{table}

\begin{table}[tb!]
    \centering
    \caption{Testbed for experiments.}
    \begin{tabular}{ll} \\\hline
        GPU    &  NVIDIA Tesla V100-PCIe\\
        Device memory &  32 GB\\
        CPU    &  Xeon Silver 4110 \\
        Host memory &  500 GB \\
        OS     &  Ubuntu 18.04 \\
        CUDA & 10.1 \\
        cuZFP & 0.5.5\\\hline
    \end{tabular}
    \label{tab:machine}
\end{table}

In this section, we provide experimental results to demonstrate the usefulness of the proposed methods in terms of reducing execution time and GPU memory consumption. 

The stencil code used in the experiments is an acoustic wave propagation program \cite{farres19eage,shen20ieice}. A 25-point stencil computation is applied to a 3-d volume that sums up to 46 GB. Such excess data was decomposed into eight chunks with 12 halo regions in conformity with 12 temporal blocking time steps. Table~\ref{tab:dataset} shows details of the stencil code. Hardware and software details of the testbed are shown in Table~\ref{tab:machine}. 

Note that the experimental results focus on the reduced execution time and GPU memory usage. The section excludes analyzing accuracy loss caused by compression, because such an analysis was given in our previous work~\cite{shen22pdcat}, showing that the accuracy loss was tolerable for our real-world stencil code after the code ran for more than 4,000 time steps.  

\subsection{Evaluation of performance improvement}
\label{subsec61}

\begin{figure}[tb!]
    \centering
    \includegraphics[width=.55\hsize]{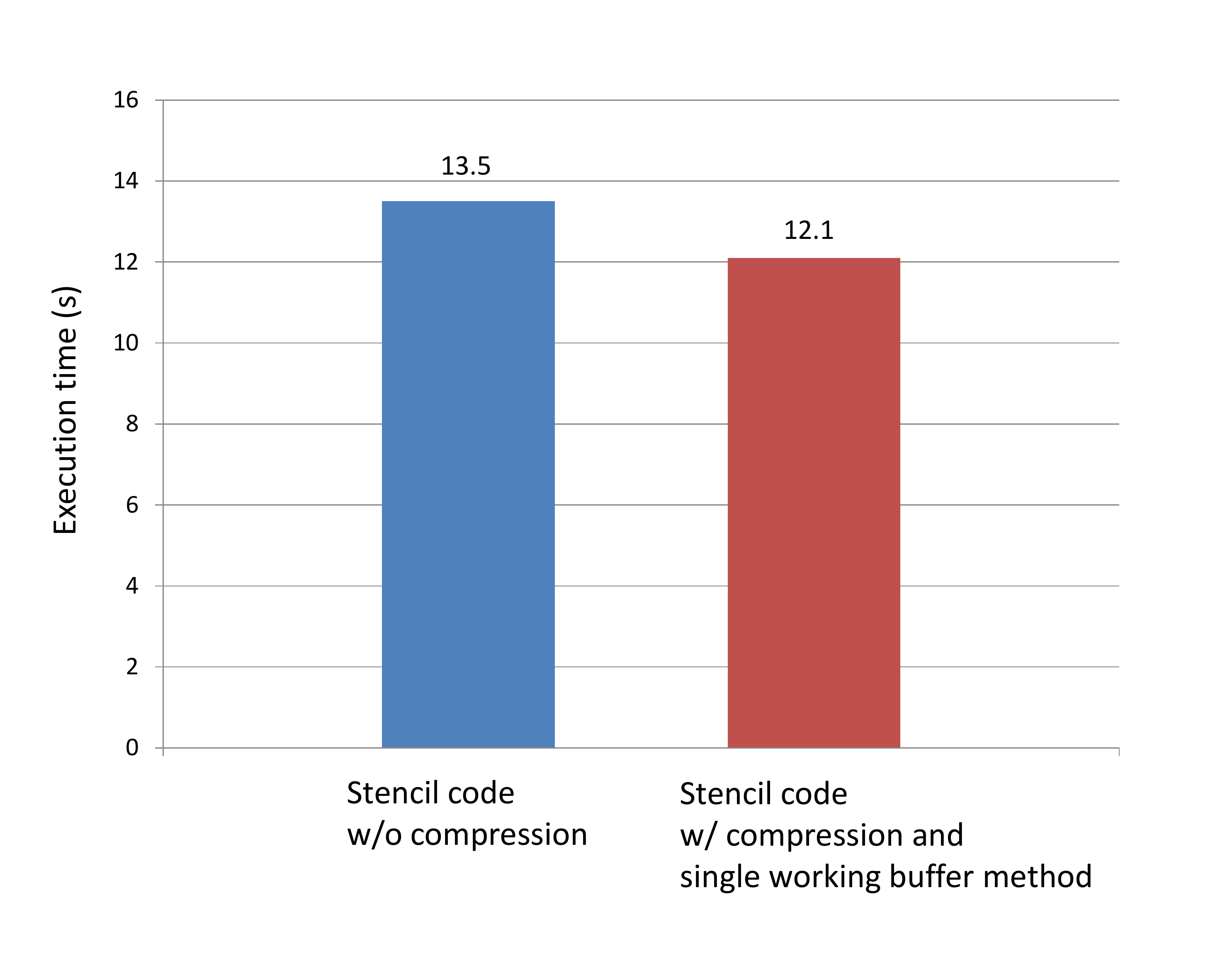}
    \caption{Comparing execution time of the code without compression with that of the code with compression and single working buffer method.}
    \label{fig:perf}
\end{figure}
\begin{figure}[tb!]
    \centering
    \includegraphics[width=.75\hsize]{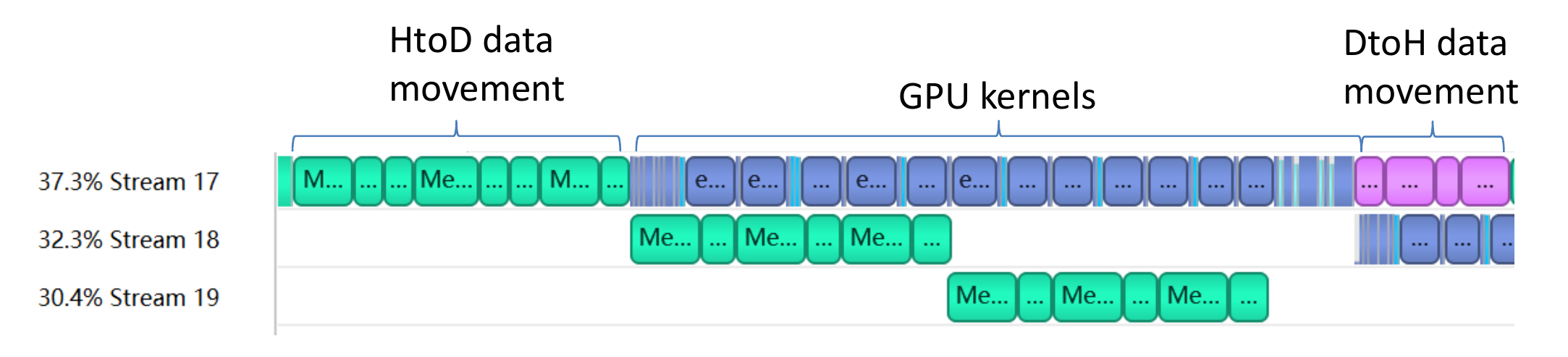}
    \caption{A part of profiling output of stencil code with proposed compression method. Note that GPU kernel time is longer than CPU-GPU data movement time, becoming the performance bottleneck.}
    \label{fig:newbot}
\end{figure}
In this section, we compare the execution time of the stencil code without compression with that of the stencil code with compression and the proposed single working buffer method. Both codes were executed for 36 time steps. 

Figure~\ref{fig:perf} shows that utilizing on-the-fly compression reduces the overall execution time of the out-of-core stencil code. The code with proposed compression method achieved a speedup of 1.1$\times$, compared to the code without compression. Such a speedup can be further increased because we did not optimize the GPU kernels of the code. Actually, the compression technique shifts the bottleneck for the stencil code from CPU-GPU data movement to GPU computation (Fig.~\ref{fig:newbot}), which is a favorable finding because optimizing GPU kernels has been much more sufficiently studied \cite{matsumura2020an5d,sai2020accelerating} than optimizing CPU-GPU data movement does.

\subsection{Evaluation of GPU memory usage improvement}
\label{subsec62}

\begin{figure}[tb!]
    \centering
    \includegraphics[width=.55\hsize]{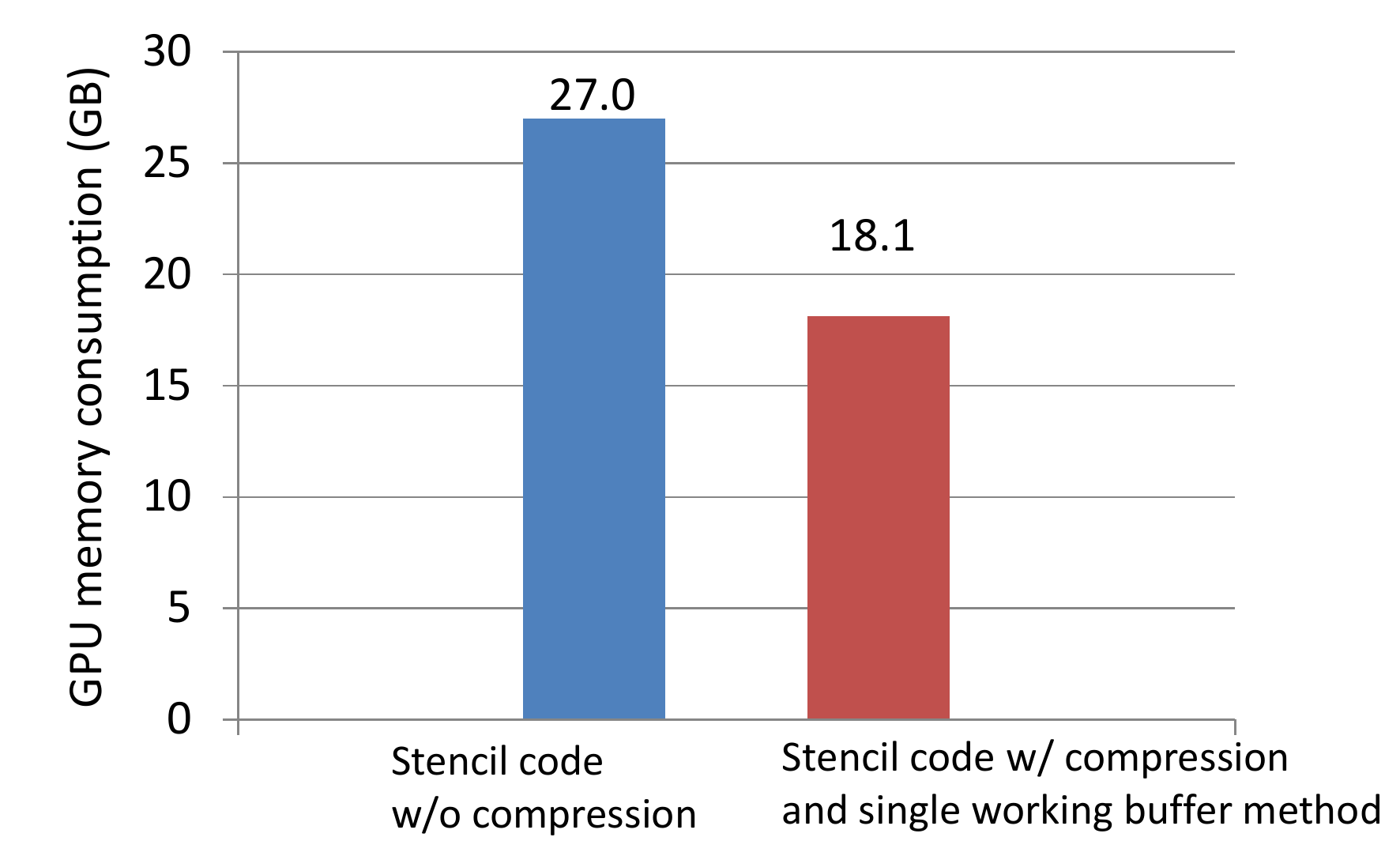}
    \caption{Comparing GPU memory consumption of the code without compression with that of the code with compression and single working buffer method.}
    \label{fig:memo}
\end{figure}
Figure~\ref{fig:memo} illustrates that the code with compression and single working buffer method significantly reduces GPU memory consumption by 33.0\%, compared to the code without compression. We can do a simple math to verify this improvement. First, the four datasets of the stencil code are a read-only dataset, two read-write datasets, and a write-only dataset. Although no data is moved to and from the write-only dataset, we still need to prepare a working buffer for it on the GPU for computation. Now assume that the size of a working buffer is 1. For the code without compression, the GPU memory consumption is 1$\times$4$\times$3$=$12, where the number 3 means three CUDA streams. 
On the other hand, for the code with compression and single working buffer method, three half-size buffers and a full-size working buffer are used, and no half-size buffer is needed for the write-only dataset. Therefore, the memory consumption is 0.5$\times$3$\times$3$+$1$\times$4$=$7.5. The ratio of the reduction is (12$-$7.5)/$12$=37.5\%. The discrepancy between the theoretical and practical results is because the procedure of on-the-fly compression consumes some additional memory at runtime.  

\section{Conclusion}
\label{sec7}
In this article, we proposed compression-based optimizations for out-of-core GPU stencil computation. On-the-fly compression was integrated into the stencil code, which improved the overall execution time by reducing memory copy (i.e., data movement) between the CPU and GPU. The performance bottleneck was thus shifted CPU-GPU memory copy to GPU computation. 
Such a shift is promising because further performance improvement can be achieved if GPU computation is optimized, which is not done in this work but many mature techniques are available. 
Furthermore, the proposed single working buffer method can significantly reduce GPU memory consumption, which improves the efficiency of resource usage.
Experimental result shows that the stencil code with proposed optimizations achieved a 1.1$\times$ speedup and saved GPU memory by 33.0\%, compared to the code without compression.

Future work includes (1) comparing various compression libraries and (2) fully releasing the potential of on-the-fly compression by optimizing GPU computation.

\backmatter


\mbox{}
\bibliography{sn-article}


\end{document}